# On the formation of low-mass black holes in massive binary stars

G.E. Brown[1], J.C. Weingartner[2]

Department of Physics, State University of New York at Stony Brook, Stony Brook, N.Y. 11794

and

Ralph A.M.J. Wijers[3]

Princeton University Observatory, Peyton Hall, Princeton, N.J. 08544-1001
and Institute of Astronomy, Madingley Road, Cambridge CB3 0HA, UK

## ABSTRACT

Recently (Brown & Bethe 1994) it was suggested that most stars with main sequence mass in the range of about $18 - 30 M_\odot$ explode, returning matter to the Galaxy, and then go into low-mass ($\geq 1.5 M_\odot$) black holes. Even more massive main-sequence stars would, presumably, chiefly go into high-mass ($\sim 10 M_\odot$) black holes. The Brown-Bethe estimates gave approximately $5 \times 10^8$ low-mass black holes in the Galaxy. A pressing question, which we attempt to answer here, is why, with the possible exception of the compact objects in SN1987A and 4U 1700–37, none of these have been seen.

We address this question in three parts. Firstly, black holes are generally "seen" only in binaries, by the accretion of matter from a companion star. High mass black holes are capable of accreting more matter than low-mass black holes, so there is a selection effect favoring them. This, in itself, would not be sufficient to show why low-mass black holes have not been seen, since neutron stars (of nearly the same mass) are seen in abundance.

Secondly, and this is our main point, the primary star in a binary —the first star to evolve— loses its hydrogen envelope by transfer of matter to the secondary and loss into space, and the resulting "naked" helium star evolves differently than a helium core, which is at least initially covered by the hydrogen envelope in a massive main-sequence star. We show that primary stars in binaries can end up as neutron stars even if their initial mass substantially exceeds the mass limit for neutron star formation from single stars ($\sim 18 M_\odot$). An example is 4U 1223–62, in which we suggest that the initial primary mass exceeded $35 M_\odot$, yet X-ray pulsations show a neutron star to be present.

---

[1] E-mail: popenoe@nuclear.physics.sunysb.edu

[2] E-mail: josephw@phoenix.princeton.edu

[3] E-mail: R.Wijers@ast.cam.ac.uk

Thirdly, we show that 4U 1700–37, the only example of a well studied high-mass X-ray binary which does not pulse, is a candidate for containing a low-mass black hole.

*Subject headings:* stars: binaries: close — stars: evolution — stars: neutron — stars: Wolf-Rayet — black holes

## 1. Introduction

For a long time it has been thought that the fact that all accurately measured neutron star masses[4] lie in a narrow band of $1.25 M_\odot < M_{NS} < 1.45 M_\odot$ is somehow a consequence of the way in which they form. Stars of main sequence mass $M > 12 M_\odot$ collapse when the Fe core reaches the Chandrasekhar limit

$$M_{CS} = 5.76 Y_e^2 M_\odot, \tag{1}$$

where $Y_e$ is the ratio of electrons to nucleons. (We omit stars of $8 - 12 M_\odot$, which are supposed to end up as neutron stars, but do not form quasi-equilibrium Fe cores in their burning and, hence, behave in a more complicated fashion.) The maximum stable mass can be increased by thermal pressure, by the factor $1 + \pi^2 T^2/\mu_e^2$, typically an approximately 15% enhancement. With $Y_{e,\text{final}} \simeq 0.43$ this gives a thermally modified Chandrasekhar gravitational mass of $\tilde{M}_{CS} \simeq 1.25 M_\odot$. The $Y_{e,\text{final}}$ is set by the strong $\beta$-decay of $^{63}$Co which opposes the electron capture proceeding to lower $Y_e$ (Aufderheide et al. 1990 and section 8.C of Bethe 1990). For stars of $12 M_\odot$, $\tilde{M}_{CS}$ will be a bit less, because the temperature is lower, whereas for heavier stars of $20 M_\odot$, $\tilde{M}_{CS}$ will be higher; but the variation is not large.

Neutron star masses can only be accurately measured when the neutron star occurs in a binary, and there are many situations in which it can accrete mass from the companion. This was generally not thought to increase the neutron star mass appreciably, because the accretion was assumed to be less than the Eddington limit, $\dot{M} \simeq 1.5 \times 10^{-8} M_\odot \, \text{yr}^{-1}$. However, it has been known for some time that if neutrinos can carry off the bulk of the energy, accretion can proceed at a much greater rate. Colgate (1971) computed the hypercritical accretion of up to $10^{-5} M_\odot$ of matter onto a neutron star. Zel'dovich, Ivanova, & Nadeshin (1972) considered the optically thick accretion of $\sim 10^{-5} \, M_\odot$ of matter over a time of 36 s. (See also Bisnovatyi–Kogan & Lamzin 1984). Chevalier (1993) pointed out that during the common envelope phase of binary evolution, photons would be trapped and accretion could occur at much higher rates, typically $10^{-4} - 10^{-3} \, M_\odot \, \text{yr}^{-1}$, and that neutron stars which have to go through this phase generally will go into black holes. Since the standard scenario for binary pulsar evolution has neutron stars going through a common envelope phase[5], in the usual situation the neutron star may have the opportunity to accrete $\geq 1 M_\odot$ of matter. Terman et al. (1994) and Taam et al. (1994) have found, in three-dimensional (non-axisymmetric) treatment of the neutron star in the common envelope, that the neutron star may survive spiral-in. This possibility arises when the neutron star ends up in a low-density region just outside a hydrogen burning shell with the massive companion in its red giant phase. By torquing up the surrounding envelope matter, it creates a hole around itself, which then leads to stopping of accretion and spiral-in. Although the authors do not calculate the accretion onto the neutron star, it is presumably less than that found by Brown (1995), although the neutron star might be expected to accrete hypercritically in the intermediate phase of common envelope evolution. Consequently, the standard scenario of binary pulsar evolution is expected to involve various gradations in the amount of matter accreted onto the neutron star. Thus, the fact that neutron stars of mass greater than the larger one of $1.44 M_\odot$ in PSR 1913+16 have not been observed might be interpreted as evidence that neutron stars heavier than this do not exist. We hope to make this point firmer in our discussion here (see our later discussion of Vela X-1 and 4U 1700–37).

---

[4]Unless otherwise indicated, quoted masses are gravitational ones.

In particular, no neutron star has been seen in SN1987A, leading Brown et al. (1992) to suggest that it went into a black hole. Chevalier (1989) has estimated that if a neutron star were present, it would be seen with a luminosity $L \sim 10^4 L_\odot$ after about a year, after which the photons from accretion of ambient matter on to it can get out. Brown & Weingartner (1994) confirmed this result. Recently, Bethe & Brown (1995) obtained an upper limit on the mass of the compact core in 1987A of $1.56 \, M_\odot$ from the $0.075 \, M_\odot$ of Fe production. We will outline their method, which will be very useful for us here, later. This upper limit was the maximum of a mass determination of $1.535 \pm 0.02 \, M_\odot$ using the presupernova core evolved by Woosley. The presupernova core of Thieleman et al. (1995) would have given about $1.443\,M_\odot$, just above the mass of the Hulse-Taylor pulsar. This latter mass may be somewhat too small because of the use of Schwarzschild, rather than Ledoux, convection in the calculations of the Nomoto group. On the other hand, as we shall discuss, using evolutionary calculations of Woosley, Langer, & Weaver (1993) together with the evolution of the Hulse-Taylor pulsar by Burrows & Woosley (1986), the pulsar mass comes out as $1.50\,M_\odot$, somewhat larger than the observed $1.44\,M_\odot$. Therefore, we believe that the upper limit on the compact core mass in 1987A is somewhat too high, and we shall adopt the estimate of $1.50\,M_\odot$ of Brown & Bethe (1994), keeping in mind that it may be wrong, in either direction, by a few percent.

---

[5] An alternative scenario is given by Brown (1995).

Having arrived at an estimate for the compact core mass of 1987A, we adopt it as the maximum possible neutron star mass,

$$M_{\rm NS,max} = 1.50 M_\odot \; . \qquad (2)$$

This argument is, of course, based on the assumption that the compact object in 1987A went into a black hole.

Brown & Bethe (1994) developed the scenario, based on the kaon condensation equation of state of dense matter (Thorsson et al. 1994), that in many cases in the collapse of massive stars the compact core is stable for a sufficient time for explosion and the return of matter to the Galaxy, and then goes into a black hole. This was estimated to happen for stars of ZAMS masses of about $18 - 30 M_\odot$, possibly including SN1987A, with progenitor mass of $18 \pm 2 M_\odot$ at the lower limit. The upper limit may exceed $30 M_\odot$ somewhat, since Swartz et al. (1993) conclude that the mass of the ejecta of the Type Ib SN1984L must have been in excess of $10 M_\odot$, and all of this in He or heavier elements. A He star of $10 M_\odot$ corresponds to a main-sequence star of about $30 M_\odot$. A similar argument can be made for SN1985F. This possibility, that a star first explodes and subsequently drops into a black hole, had been suggested by Wilson et al. (1986) and Woosley & Weaver (1986). They had in mind the conventional scenario of a neutron star in which thermal pressure and neutrino pressure stabilize the compact object during the Kelvin-Helmholtz contraction and then later, as heat is carried away by neutrinos, the core collapses into a black hole. Prakash et al. (1995) show by detailed calculation that this is indeed possible, but only for a small interval of $\Delta M \sim 0.05 - 0.1 M_\odot$ in compact core masses, the upper limit depending on the somewhat unknown late time fall back of material. In addition to this "window" from thermal pressure and late time fallback, Brown & Bethe (1994) find approximately an additional $0.2 M_\odot$, i.e., a total window of $\Delta M = 0.25 - 0.3 M_\odot$, from the properties of the kaon condensed EOS. Chiefly this results because, at high densities, the matter ends up as nuclear matter, not neutron matter. The former is much "softer" than the latter, and sends the core into a black hole.

Although the Brown & Bethe (1994) scenario indicated that most single stars with main-sequence masses between 18 and $30 M_\odot$ exploded, returning matter to the Galaxy, and left low-mass black holes, the Fe cores do not increase monotonically with main-sequence mass (Timmes, Woosley, & Weaver 1995) and this complicates the situation. In Table 1 we list their baryon number masses, together with the gravitational masses obtained from the Lattimer & Yahil (1989) binding energy correction

$$E = 0.084 M_\odot \left(\frac{M}{M_\odot}\right)^2, \qquad (3)$$

where $M$ is the gravitational mass of the compact core. We shall later outline a way of estimating the bifurcation radius, which, we believe, leads to an improved estimate of the final compact core mass. Since supernova explosions giving quantitative results have not yet been carried out, Timmes et al. (1995) chose mass cuts (Table 1) outside the neutronized iron core and at the location of an abrupt entropy jump if one were nearby. An unknown amount of material will accrete onto the compact core as the explosion develops. We should also note that their work shows substantial changes in compact core masses with metallicity.

The suggestion of low-mass black holes was somewhat new (although, as noted above, it had been made earlier) and the number of about $5 \times 10^8$ in the Galaxy estimated by Brown & Bethe naturally raised the question as to why none had been seen. (The compact core of SN1987A has not been observed.) Part of the reason is that they would be most visible in high-mass X-ray binaries (HMXB's) and these last only $2 - 5 \times 10^4$ years (Van den Heuvel 1976). There are, however, at least four good candidates for high-mass black holes, and there should be roughly as many in which the compact object is a low-mass black hole (Brown & Bethe 1994).

One would expect more high-mass black holes to be seen, since the accretion luminosity from Bondi-Hoyle type wind accretion increases with the square of the mass of the accreting object up to the Eddington luminosity. Above that, the luminosity is limited to the Eddington rate, which increases linearly with mass. High mass black holes with $M \sim 10 M_\odot$ can thus give about two orders of magnitude more accretion than the low-mass ones with $M \sim 1.5 M_\odot$. Hence, the larger number of high-mass black holes observed may be somewhat of a selection effect, relative to the low-mass black holes.

In Fig. 1 and Table 3 we show the known masses of compact objects. The masses of radio pulsars, at the bottom of Fig. 1, all fit nicely in with our estimate of eq. 2 for $M_{\rm NS,max}$. Error bars in the masses of the high-mass X-ray binaries are large. It is nonetheless interesting to discuss 4U 1700–37 and Vela X-1. Not only do the central values of their masses exceed our $M_{\rm NS,max}$ the most (In fact, they are the only cases where the central values do exceed our $M_{\rm NS,max}$.), but these two X-ray binaries are the closest in distance, both lying within 3 kpc.

Van Kerkwijk et al. (1995b) find that observed velocities in Vela X-1 deviate substantially from the smooth radial-velocity curve expected for pure Keplerian motion. The deviations seem to be correlated with each other within one night, but not from one night to the other. The excursions suggest something like pulsational coupling to the radial motion and make it difficult to obtain an accurate mass measurement. The lower limit for the mass of the compact object in Vela X-1 is now found to be $1.43 M_\odot$ at 95% confidence lower limit or $1.37 M_\odot$ at 95% confidence interval around the most probable value (Van Kerkwijk et al. 1995b). Because of the pulses in the X-ray spectrum, the compact object in Vela X-1 is known to be a neutron star.

It is interesting to speculate on other possibilities for low-mass black holes in binaries. We suggest that the compact object in 4U 1700–37 is a candidate. Heap & Corcoran found a mass of the compact object of

$$M_C = 1.8 \pm 0.4 M_\odot. \qquad (4)$$

Of course, the errors are large, and in addition there are systematic problems. Contrary to all other high-mass systems with such a low-mass compact star, it is not pulsing in X rays (Bhattacharya & Van den Heuvel 1991), so it does not appear like a rotating magnetized neutron star. We shall expand our discussion of 4U 1700–37 in the next section.

The Brown & Bethe (1994) scenario was painted with broad strokes. Whereas the chief points may be correct, individual events have special features, such as the fluctuation of Fe cores with main-sequence mass. Even though quantitative calculations of the entire supernova have not been carried out to date, and it may take some time until accurate ones are completed, it is interesting to try to correlate observations with the general picture of Brown & Bethe (1994).

It will become clear in our discussion that SN1987A was a unique event, allowing us to observe what was going on (or not going on) in a single star, whereas all other mass determinations for compact objects are in binaries. During the evolution of a binary, the initially more massive star, which we call the primary, will evolve first, transferring mass to the secondary (which then may become heavier than the remaining primary). The primary is thought to be left, in most cases, without hydrogen envelope: it is a helium, or Wolf-Rayet, star. This helium star does not, however, evolve like the helium core of the original main-sequence star with hydrogen envelope. The core evolution and nucleosynthesis are altered if substantial mass loss continues, as it usually does, after the helium core is uncovered (Woosley, Langer, & Weaver 1993). Large mass loss is likely, according to these authors, to lead to final helium star masses as small as $4 M_\odot$ for a wide range of initial masses, such as the $35 - 85 M_\odot$ range studied. This occurs because the mass loss rate is mass-dependent. Simply integrating their mass loss formula,

$$\dot{M}_{\rm WR} = 5 \times 10^{-8} \left(\frac{M_{\rm WR}}{M_\odot}\right)^{2.6} {\rm M}_\odot \, {\rm yr}^{-1}, \tag{5}$$

over $10^6$ yr we find that 20, 10, and $4\,M_\odot$ helium stars end up at 4.6, 4.1, and $2.8\,M_\odot$, respectively. These numbers are not far from those arrived at by the full evolution calculation, so it is clear that the final masses are almost completely determined by the mass loss rate $\dot{M}_{\rm WR}$. This result cannot yet be considered very well established, because measurements of masses and mass loss rates are usually quite uncertain. The available data sometimes yield a much shallower dependence of mass loss rate on mass, in which case the strong mass convergence noted here does not occur (see, e.g., Langer 1989, Schmutz, Hamann, & Wessolowski 1989, Smith & Maeder 1989). We shall nevertheless stick to this mass loss prescription, since detailed calculations are available for it.

The chief result of Woosley, Langer, & Weaver (1993) is that a presupernova star is not uniquely specified by its initial helium core mass. The presupernova star carries a memory, especially in the size of its CO core and its surface composition, of its earlier evolution. In order to show why a naked helium star ends up with a smaller Fe core mass than an initially "covered" helium core of the same mass, which resulted by loss of mass by wind from a massive main-sequence star, they evolve a $4.25 M_\odot$ naked helium core and a $4.25 M_\odot$ helium core that resulted after mass loss by wind from a $60 M_\odot$ main-sequence star. Their chief point is that the latter core retains a "chemical memory" (although not a "thermal memory") of its earlier history when it was covered up. The convective core size at the end of helium core burning is similar ($M_{\rm CC} \simeq 2 M_\odot$) in the two cases. However, the chemical composition just outside this core is very different. In the case of the initially covered core, most of the matter just above the convective core has been burned to carbon and oxygen (presumably the "wraps" have kept the region hotter) so that there is very little helium left. In the case of the naked helium core, the helium concentration rises to 100% immediately beyond the convective zone. In the case of the initially covered helium core, the helium burning shell which develops at core helium exhaustion moves rapidly outwards, through the small helium concentration, but for the naked star with $Y \simeq 1$, it remains almost fixed in mass at the edge of the convective core. Consequently, the carbon-oxygen core masses of the presupernova models are very different in the two cases, $3.03 M_\odot$ for the initially covered case and $2.12 M_\odot$ for the always naked case. This leads to a smaller Fe core for the naked case, and a better chance that it will end up as a neutron star.

A "naked" helium core is left after mass transfer from the primary in a binary, and such helium cores have large mass loss (eq. 5). The iron core mass resulting from the burning of these "naked" helium cores cannot be inferred from the main-sequence mass because once mass loss eats substantially into the He core it changes the CO core mass as noted above. Woosley, Langer, & Weaver (1994) find that the composition of one of these stars is different from that of any star evolved without mass loss (see also Woosley et al. 1993). It would be identical with that of helium cores in main-sequence stars if no mass were lost. The iron cores of Woosley, Langer, & Weaver (1994) are given below in Table 2. Note that these iron core masses are substantially less than those given in Table 1, where the cores were evolved with hydrogen envelope present. For example, the $1.49 M_\odot$ Fe core for a $10 M_\odot$ helium star should be compared with the $1.99 M_\odot$ Fe core for a $30 M_\odot$ main-sequence star in Table 1. Consequently, we see that "naked" primaries in binaries are much more likely to end up as neutron stars than single stars of the same main-sequence mass.

The Fe core mass will not give the entire mass of the compact object, as there will be fallback from out to the bifurcation radius. Following Thielemann et al. (1990) and Bethe (1990) we can estimate this radius from the fact that a small amount, about $0.075 M_\odot$, of Fe came off from SN1987A. This means that bifurcation had to come at a radius slightly inside of that up to which oxygen and silicon were burned to $^{56}$Ni, which later went into Fe through weak decays. Corrections to these estimates have been considered in detail by Bethe & Brown (1995). However, they all essentially cancel out, leaving the simple picture we outline here.

To form $^{56}$Ni from $^{28}$Si by successive addition of $\alpha$ particles, the temperature must be above $T = 350\,\text{keV} = 4 \times 10^9\,\text{K}$. Given that the energy is mostly in radiation and electron pairs, $T = 350\,\text{keV}$ corresponds to a black body energy density of

$$w = 3.5 \times 10^{24}\,\text{ergs cm}^{-3}. \tag{6}$$

The shocked system is, to a good approximation, isothermal, so the energy density is also

$$w = \frac{E}{4\pi R^3/3}, \tag{7}$$

where $R$ is the shock radius. It is then straightforward to find that

$$R = (4100\,\text{km}) E_{51}^{1/3}, \tag{8}$$

where $E_{51}$ is the total energy in foes. Estimates for SN1987A give $E_{51}$ in the range of $1 - 1.5$; therefore

$$4100\,\text{km} < R < 4700\,\text{km}. \tag{9}$$

The detailed calculations of Bethe & Brown (1995), including corrections, give $R = 3900 \pm 400\,\text{km}$, not very different from (9). Hence, it may be reasonable to choose the enclosed mass somewhere in this range as the mass which will end up in the compact core. The $M_{3500}$ and $M_{4500}$ in Table 2 were kindly furnished us by Stan Woosley (1994).

In fact, Woosley, Langer, & Weaver (1994) find that for 10 explosions of Wolf-Rayet stars of various masses in the range of $4 - 20 M_\odot$, the mass of $^{56}$Ni is small, lying in the narrow range $0.07 - 0.15 M_\odot$. In a fit to the light curve of the recent Type Ic supernova 1994I a $^{56}$Ni mass of $0.04 - 0.07 M_\odot$ is derived. Thus, our procedure of obtaining the bifurcation radius near the edge of the iron core, as was done in SN1987A, finds support. Supernova explosion energies in the range of those noted above for SN1987A resulted, although in one case the energy was increased up to $E_{51} = 1.7$ in order to obtain an improved agreement with the light curve of SN1994I. A $5 M_\odot$ helium star was exploded in this case.

Swartz et al. (1993) specify that SN1987M, a nominal type Ic supernova, must have exploded with a helium surface abundance near 7%. Even the massive $20 M_\odot$ helium star of Woosley, Langer & Weaver (1994), which corresponds to a main-sequence star of $45 - 50 M_\odot$, has 33% helium at the surface and 12% when averaged over all ejecta. This suggests to us that the mass loss rates employed by Woosley, Langer, & Weaver (1994) are not too large. Of course, SN1987M might have come from an even more massive star, in which case the helium abundance would have been less, but there are not many such very massive stars. In any case, very large mass loss rates in Wolf-Rayet stars are necessary in order to account for Type Ic supernovae, if these are to be formed from single stars. Note, however, that there is an alternative scenario (Nomoto et al. 1994) in which spiral-in of a compact object rather than a wind is used to dispose of the envelope and a strong wind is therefore not needed.

In Table 2 we have included $M_{3500}$ and $M_{4500}$, the mass for the Wolf-Rayet cores enclosed by radii at 3500 and 4500 km. It should be noted that these enclosed masses vary slowly and smoothly with initial helium core mass, over a wide range of corresponding main-sequence masses.

We shall consider the two high-mass X-ray binaries 4U 1700–37 and 4U 1223–62 in the next section. They have the largest mass functions of the dozen or so high-mass X-ray binaries for which the orbital parameters have been determined. As noted, the central value of the mass of the compact object in 1700–37 is large, and it does not pulse. From these considerations it seems quite possible that the compact object is a low-mass black hole. The binary system 4U 1223–62 is thought to have a companion star of about $50\ M_\odot$. X-ray pulsations with a period of 11.6 minutes were discovered in this binary by White et al. (1976), so it is known to contain a neutron star. It is a major challenge to see how the primary could have ended up as a neutron star.

In the next section we shall show, from evolutionary arguments, that it is reasonable that 1700–37 is the only one of the well measured high-mass X-ray binaries to (probably) contain a low-mass black hole.

## 2. The high-mass X-ray binaries 4U 1700–37 and 4U 1223–62

The companion star HD 153919 in 4U 1700–37 is an O6f star. Based on orbital solutions reviewed by Hutchings (1976), Conti (1978) chose a value of $27 M_\odot$ for this star, although he noted that the value was uncertain. Consistency with stellar evolution could not be found for this mass, however, and Conti noted that "HD 153919 is either overluminous or undermassive...", as were all five of the companion stars of massive X-ray binaries he investigated. According to Ziółkowski (1979, and references therein) this is what one expects: massive stars evolve at roughly constant luminosity, but lose a significant fraction of their mass due to stellar wind even on the main sequence. This loss hardly affects their core structure, and thus also hardly affects their luminosity. The luminosity therefore reflects its initial mass, whereas of course the binary motion measures its current mass. Ziółkowski estimates that a mass loss of about $10^{-5.5}\,M_\odot\,\mathrm{yr}^{-1}$ can account for the observed discrepancy between mass and luminosity; such a value is right in the range of measured mass loss rates of massive O stars (see, e.g., Lamers & Leitherer 1993). Note that this mass loss increases the ratio of core mass to total mass of the star, making it look like a giant (some of them are indeed classified as such from their spectra) while in fact it is still on the main sequence.

Recently Heap & Corcoran (1992) claim to have resolved the 'problem' of Conti, deriving a mass for the O6f star of $52 \pm 2 M_\odot$. These authors show how the optical thickness of the wind to low energy X rays can cause an overestimate of the duration of the X-ray eclipse and hence, an underestimate of the O star mass. They derive a mass which is nearly double the mass derived by Conti and in fair agreement with the mass of the star deduced from its spectroscopic properties and location on the H-R diagram. However, most of the remaining problems noted by Heap & Corcoran (1992) at the end of their paper could be most easily resolved by lowering the mass of HD 153919 somewhat. The problem is that their paper relies heavily on knowing what the properties of the wind of a star with given mass ought to be by comparison with single stars, whereas it is entirely unclear that the companion, almost filling its Roche lobe, should have wind properties similar to that of a single star of the same mass. The situation is unsatisfactory, and we settle on an estimate of $40 \pm 10 M_\odot$ with error bars large enough to encompass both of the above estimates. Indeed, Heap & Corcoran say that HD 153919 is much like $\lambda$ Cep. Herrero (1995) finds a mass for $\lambda$ Cep in the central part of this range.

From measurements of the orbital parameters in 1223–62, Sato et al. (1986) determined the mass of the companion, Wray 977, to be $M_{\rm opt} \geq 33 M_\odot$. From the lack of X-ray eclipses, the limit on inclination angle is estimated to be $i \leq 78°$. This increases the lower limit on the companion mass to be $M_{\rm opt} \geq 35 M_\odot$. They find a plausible set of parameters to be $i \simeq 75°$ and $M_{\rm opt} \simeq 38 M_\odot$, which takes into account all X-ray and optical constraints. More recently, Kaper et al. (1994) revised the spectral classification of Wray 977, claiming it is a hypergiant, and thus further away from us. This more than doubles the star's radius and thus forces a smaller inclination ($i \leq 62°$) in order to avoid eclipses. Consequently, the minimum mass is 48 $M_\odot$.

It seems likely that the primary, which exploded, was also heavy. Single stars in the mass range $M \gtrsim 40 M_\odot$ rapidly lose their hydrogen–rich envelopes and turn into Wolf–Rayet stars. The absence or presence of a companion makes no difference (Chiosi & Maeder 1986). Woosley et al. (1993) find that rapid mass loss in the luminous blue variable phase determines the stellar mass at the beginning of helium burning. The hydrogen–rich envelope is completely gone. There is, thus, a situation similar to that of the "naked" helium stars, although the helium core retains a "chemical memory" of its evolution, as noted above.

Woosley et al. (1993) find that the helium core mass and further outcome is strongly influenced by the $^{12}\rm{C}(\alpha,\gamma)^{16}\rm{O}$ rate, which is of key importance for central helium burning and all further burning stages. For a value of this rate somewhat larger than that favored by stellar nucleosynthesis, a $60 M_\odot$ star develops a helium core mass of only $4.25 M_\odot$ (as discussed above) and a baryon number iron core mass of $M_{\rm Fe} = 1.40 M_\odot$. For a $^{12}\rm{C}(\alpha,\gamma)^{16}\rm{O}$ rate somewhat smaller than favored by nucleosynthesis, the results are $M_{\rm He} = 6.65 M_\odot$ and $M_{\rm Fe} = 1.46 M_\odot$, somewhat larger. We note that these are not very different from the Fe core masses for helium cores of the same mass in Table 2. It should be said that the $M_{3500}$ and $M_{4500}$, which we give below, for the $60 M_\odot$ stars are somewhat larger than those in Table 2 for the same mass helium core, indicating a somewhat different entropy profile. Gravitational masses for the compact core of the $60 M_\odot$ star with large $^{12}\rm{C}(\alpha,\gamma)^{16}\rm{O}$ rate are $M_{3500} = 1.48 M_\odot$ and $M_{4500} = 1.53 M_\odot$, whereas for the small rate they are $M_{3500} = 1.57 M_\odot$ and $M_{4500} = 1.66 M_\odot$. The compact core masses for the $40 M_\odot$ star evolved by Woosley et al. (1993) are similar to those for the $60 M_\odot$ one, so there is presumably little difference in the region of masses $40 - 60 M_\odot$. In fact, Woosley et al. note that all of their models, with the exception of the heavier $85 M_\odot$ one, have strikingly similar iron cores, cores that are also similar in mass to lighter presupernova stars arising in the 12 to $35 M_\odot$ range. And they note "Thus it seems likely that whatever mechanism functions to explode the common Type II supernova will also operate for at least some of these stars. There is no apparent mass limit above which one can say that a black hole mass remnant is very probable."

Given the very many uncertainties in the evolution of heavy stars with mass loss, it might appear unreasonable to consider the fact that most of the $M_{3500}$ and $M_{4500}$ masses exceed the Brown–Bethe $1.50 M_\odot$ limit for neutron star masses. However, given the Bethe & Brown (1995) determination, as an exercise, we will do just that. Woosley et al. (1994) have found large effects, due to mass loss by mass transfer, in the evolution of helium cores, the "naked" ones shown in Table 2. We suggest that the presence of the companion hastens mass loss sufficiently in the heavy stars of mass $40 - 60 M_\odot$ in order to bring the gravitational masses of their compact cores in at least some cases below $1.5 M_\odot$. After all, the effective gravity of the primary is lowered by the presence of the companion star, and this is especially so when the star almost fills its Roche lobe. Only small effects from the presence of the companion are required to bring the companion mass down to $\leq 1.5 M_\odot$. The importance of such small changes as we require (desire) would not have been appreciated previously.

We now briefly consider the relation between the current mass of the optical companion in a high-mass X-ray binary and the mass of the initial primary when the binary was on the zero-age main sequence. The reason is that we wish to estimate the progenitor masses of the neutron stars in observed X-ray binaries in order to empirically establish the mass of a main-sequence star in a close binary that may leave a neutron star remnant. It does of course not suffice to simply use the current mass of the optical companion for this purpose, since that is an evolved star which has gained mass from the initial primary. We follow the work of Van den Heuvel & Habets (1984), with small modifications. We use the evolution tracks of Maeder (1990), since he used the same mass loss prescription as in the work of Woosley and collaborators discussed above. A high-mass X-ray binary as observed now is assumed to have started out as a close binary with primary mass $M_p$ and mass ratio $q$. When the primary reaches the end of the main sequence, it expands and transfers its hydrogen envelope to its companion (so-called case B mass transfer); a fraction $f$ of the transferred mass is lost from the binary. Then the now naked helium star primary evolves rapidly to a supernova and explodes. We neglect the short time this takes. Meanwhile, the now more massive secondary is evolved to the end of its main-sequence life. We accounted for rejuvenation by the added mass (Van den Heuvel 1969) when computing the time from mass transfer to core hydrogen exhaustion in the secondary. The occurence of rejuvenation was recently questioned by Braun & Langer (1995) in certain cases. It makes little difference to us, since all we need it for is to compute the amount of wind mass loss in this phase, which is not large compared to uncertainties in our understanding of mass transfer anyway. We assume that core hydrogen exhaustion in the secondary marks the start of the X-ray binary phase. This is reasonable because the observed high-mass X-ray binaries are in fairly close binaries, and the expansion of a star from the end of the main sequence across the Herzsprung gap is fast, so not much time will pass after the end of the main sequence until substantial accretion starts. The value of $q$ is unknown, of course, and the value of $f$ is rather uncertain: while mass transfer between roughly equal-mass stars is often thought to be conservative, there are indications that it may not be in practice, especially if the donor is a giant. We will vary these unknown parameters to estimate their importance.

As an example, consider a binary with initial masses 45 and $36\,M_\odot$ ($q \simeq 0.8$). When the primary reaches TAMS (terminal age main sequence), wind losses have reduced the masses to 40 and $33\,M_\odot$. The $22\,M_\odot$ envelope of the primary is now transferred, during which 20% (say) is lost from the system. Now the stars are 18 and $51\,M_\odot$, and soon thereafter the $18\,M_\odot$ helium star explodes, leaving a $1.5\,M_\odot$ compact object. When the rejuvenated secondary reaches TAMS, wind losses have reduced it to $48\,M_\odot$, implying that in this case an X-ray binary has formed with an optical companion of mass $48\,M_\odot$, the lowest allowed value for 4U 1223–62. Inspection of Table 2 shows that an $18\,M_\odot$ helium star evolves to a core with a gravitational mass of just above $1.5\,M_\odot$. Given the uncertainties in the calculations, one may therefore state that the Brown-Bethe theory is consistent with the appearance of a neutron star companion in 1223–62.

The lowest possible value for the initial primary mass (given a target value for the eventual optical companion mass) is obtained by maximizing the initial mass of the binary, $M_\mathrm{p}(1+q)$, and minimizing mass loss from the system, i.e., setting $q=1$ and $f=0$. For optical companion masses of 48, 40, and $30\,M_\odot$ we thus find minimum required initial primary masses of 34, 27, and $19\,M_\odot$. Now consider 4U 1223–62 again: the current optical companion is at least $48\,M_\odot$, hence the initial primary must have been at least $34\,M_\odot$. It left a neutron star, thus we conclude that stars of about $35\,M_\odot$ can leave a neutron star if they evolve in a close binary. More realistic values of $f$ and $q$ could be 0.2 and 0.8. For those adopted parameters, the neutron star in 1223–62 had a progenitor of at least $45\,M_\odot$. Such massive stars may be luminous blue variables, which can become Wolf-Rayet stars without being helped by extra mass loss due to transfer in a binary, and according to Woosley et al. they would form small enough cores to lead to neutron star formation. Although it seems possible to find primary masses for which 1700–37 could go into a low-mass black hole, our considerations show that a primary in a close binary may practically never do so.

## 3. Discussion

We have discussed the scenario in which the primary star in a binary, as long as its mass is less than about $40 M_\odot$, will evolve quite differently from an isolated star of the same mass, because of transfer of its hydrogen envelope to the secondary. In this way primaries corresponding to main-sequence masses as massive as $35 - 45 M_\odot$ can evolve into neutron stars, whereas single stars in this mass range would go into low-mass black holes. The different behavior of "naked" helium cores goes, at least some way, towards explaining why only one possible low-mass black hole has been observed in high mass X–ray binaries. Indeed, even in the case of 4U 1700–37, research workers had considered the compact object here to be a neutron star with a weak enough magnetic field so that no X-ray pulsations are observed. Such an explanation for the nonpulsing of 1700–37 seemed reasonable when research workers believed in field decay of the magnetic fields of neutron stars. Recently, Taam & Van den Heuvel (1991) have shown that empirically field decay is inversely correlated with mass accretion (although there is up to now no fundamental theoretical basis for this correlation). Many examples are produced for lack of field decay in isolated pulsars.

The empirical relationship between field decay and accretion has been modelled by Shibazaki et al. (1989) by
$$B/B_0 = \frac{1}{1 + \Delta M/m_B}, \qquad (10)$$
where $B$ and $B_0$ are the current and initial magnetic fields, $\Delta M$ is the amount of matter accreted, and $m_B$ is a constant. Given that millisecond pulsars are thought to have accreted 0.01-0.1 $M_\odot$ of material and thereby decreased their magnetic field by 4 orders of magnitude, one finds $m_B \sim 10^{-5.5} - 10^{-4.5}$ $M_\odot$. 1700–37 may have been accreting material for up to $5 \times 10^4$ yr, at a rate of perhaps 10% of the Eddington rate, implying $\Delta M \lesssim 10^{-3.7}$ $M_\odot$. This means that its field could have decayed to a few percent of its initial value, possibly putting it at about $3 \times 10^{10} - 3 \times 10^{11}$ G now. The lower end of this range is marginally enough to not see pulsations, so a judgment of whether the absence of pulsations is consistent with the presence of a neutron star in this system will have to await the coming of better models for field decay in accreting neutron stars.

The number ratio of massive stars that form low-mass black holes to those that form neutron stars is low even if all stars were single, typically less than about 20% for reasonable slopes of the initial mass function. The fact that a significant fraction of O and B stars are in close binaries will lower this ratio, as we have shown. This implies that it is not at all unlikely to find no low-mass black holes among the 10 well-studied X-ray binaries in globular clusters. Hence the absence of known low-mass black holes in the population of globular-cluster X-ray sources is quite consistent with the Brown-Bethe scenario for low-mass black hole formation, and statements to the contrary by Kulkarni, Hut, & McMillan (1993) are incorrect.

It is interesting that Burrows & Woosley (1986) find, taking mass loss from the helium star and possible kick velocities into account, that the most reasonable evolutionary scenario for PSR 1913+16 begins with He stars of initial masses about $7 M_\odot$. We note, however, that the average gravitational mass of $M_{3500}$ and $M_{4500}$ for a helium star initial mass of $7\,M_\odot$ is $1.50\,M_\odot$ (Table 2). This is just $0.06\,M_\odot$ greater than the $1.44\,M_\odot$ of the Hulse-Taylor pulsar. Thus, we suspect that the Woosley et al. (1994) compact objects are somewhat too large, as noted in sect. 1. Alternatively, if we use the Woosley et al. masses, then we should use the maximum neutron star mass of $1.56\,M_\odot$ of Bethe & Brown (1995), which was derived using the Woosley presupernova core. From Table 2 we see that only $M_{4500}$ for initial He star masses in the range $10$–$20\,M_\odot$ exceeds this limit. Furthermore, from the Woosley et al. (1993) results quoted earlier, we see that their single $60\,M_\odot$ ZAMS star may or may not give a compact core mass exceeding this limit, depending on the $^{12}C(\alpha,\gamma)^{16}O$ rate. As noted earlier, the presence of a companion may help to bring the compact core mass below the maximum neutron star mass.

From these estimates, based on the results of Woosley et al. (1993), we might expect stars in binaries with helium cores up to $10\,M_\odot$ (corresponding to $30\,M_\odot$ ZAMS stars) to end up as neutron stars. In addition, luminous blue variables in the ZAMS mass range $40$–$60\,M_\odot$ may end up as neutron stars, especially in close binaries. This leaves only a narrow region of masses around $30$–$40\,M_\odot$ for possible evolution into low-mass black holes, in addition, possibly, to some very massive stars above $60\,M_\odot$. Although several aspects of our discussion are uncertain, it does seem clear that few stars in binaries would be expected to go into low-mass black holes.

In any case, it is clear that stars of main-sequence mass substantially heavier than the $18 M_\odot$ progenitor of 1987A can, if uncovered by transfer of their hydrogen envelope in binaries, end up as neutron stars, whereas the "covered" core in SN1987A probably went into a low mass black hole (Brown & Bethe 1994). Arguments were given by Brown & Bethe (1994) following those of Maeder (1992, 1993) from the relative abundance of produced helium to metals (i.e. $\Delta Y/\Delta Z$) that stars above a main-sequence cutoff mass of

$$M_{\text{cutoff}} = (25 \pm 5) M_\odot \tag{11}$$

had to go into (presumably high-mass) black holes without exploding and returning matter to the Galaxy. Given the developments of Woosley et al. (1993, 1994) outlined in this paper, this estimate will have to be revised.

Clearly the original main-sequence mass of naked helium stars in binaries which go into high-mass black holes will be much higher than this. The estimate of Brown & Bethe (1994) of the dividing line at gravitational mass of $1.84 M_\odot$, including binding energy correction from eq (3) of $0.25 M_\odot$, would correspond to a baryon number mass of $2.09 M_\odot$. This number is somewhat uncertain, since the thermal pressure was only roughly estimated, etc., but we can say that for baryon number mass $\geq 2 M_\odot$ evolution into a high-mass black hole is likely. In Woosley et al. (1993) only their highest mass star ($85 \, M_\odot$) satisfies this.

The standard scenario for black hole production is summarized by Van den Heuvel (1994). In his figure 75, a possible model for the formation of A0620–00, a low-mass X-ray binary containing a high-mass black hole (of several solar masses) is sketched, following the scenario of De Kool et al. (1987). In this scenario a $40 M_\odot$ main-sequence star loses mass by wind and mass transfer to a low-mass companion star, leaving a $12 M_\odot$ helium star. After further wind loss, the helium star drops into a black hole of $8 M_\odot$. Note from our Table 2, however, that for a helium star mass of $12 M_\odot$, the $M_{4500}$ is $1.74 M_\odot$ (baryon number mass). According to the Brown & Bethe (1994) scenario, such a star would first explode, returning matter to the Galaxy, and then go into a low-mass black hole, of mass not much larger than $1.5 M_\odot$. In other words, it would have the same fate as the one we outlined above as probable for 1700–37. Still, four good candidates for high-mass black holes are listed by Van den Heuvel (1992). While our considerations may make their formation somewhat more difficult, there are a number of known very massive WR stars (Cherepashchuk, 1991, and references therein) of which it is hard to imagine that they would not form massive black holes.

Our chief point, based on the evolutionary calculations of "naked" helium stars by Woosley et al. Weaver (1994), is that such stars, corresponding to main-sequence masses of up to $35 - 45 M_\odot$, can explode and go into neutron stars. Thus, primary stars in binaries can have a substantially greater mass than the limiting main-sequence mass estimated by Brown & Bethe (1994) and still end up as neutron stars. We suggest that it is also possible that stars in the main-sequence mass range $40 - 60 M_\odot$ may, because of rapid mass loss and because of a companion, end up as neutron stars. Indeed, Woosley et al. (1993) find that the relation between initial main–sequence mass and the mass of the presupernova star are nonmonotonic and have a maximum close to the critical ZAMS mass for WR formation, $\sim 35 - 40 \, M_\odot$. This region seems to offer the best chance, in the range of masses considered here, for low mass black hole formation. We also suggest here that it is the presence of the companion which may tip the balance in favor of the neutron star, rather than the low-mass black hole. We showed that 4U 1700–37, the only example of a well studied X-ray binary which does not pulse, is a fair candidate for containing a low-mass black hole. We conclude that neutron stars will result from many stars in binaries which, without a companion, would go into low-mass black holes. Thus, for single stars, the Brown–Bethe scenario in which stars of main-sequence masses $18 - 30 M_\odot$ explode, returning matter to the galaxy, and then go into low mass black holes, may be roughly correct.

A possibly very useful by-product of our investigation is to note the smoothness in the change of the enclosed masses $M_{3500}$ or $M_{4500}$ with either ZAMS mass or He core mass. Bethe & Brown found the bifurcation radius of $3900 \pm 400$ km for 1987A and, as noted, Woosley et al. find that in essentially all of their explosions the amount of Fe produced is similar to that in 1987A. Since this small amount of Fe was the basis for the Bethe & Brown determination, the mass enclosed at some radius between 3500 and 4500 km is our best estimate of the compact core mass. Even though the explosion mechanism for type II supernovae seems now to be understood, we believe that it will be some time before a more accurate way of estimating the compact core mass will be found.

We would like to thank Ed van den Heuvel for suggesting that 1700–37 contained a black hole. We are extremely grateful to Stan Woosley for sending us nearly all of the evolutionary material on which our arguments are based and for the many private communications, some of which are quoted in the paper. We are grateful to Hans Bethe for helpful criticism. G.B. and J.W. are supported by the U.S. Dept. of Energy Grant DE-FG02-88ER 40 388; R.W. is supported by a Compton Fellowship (grant GRO/PFP-91-26) and by a PPARC fellowship.

Table 1: Compact core masses for solar metallicity, from Timmes, Woosley, & Weaver (1995)

| Main Sequence Mass/$M_\odot$ | Baryon Number Mass/$M_\odot$ | Gravitational Mass/$M_\odot$ |
|---|---|---|
| 15 | 1.57 | 1.40 |
| 18 | 1.64 | 1.46 |
| 20 | 1.54 | 1.38 |
| 25 | 1.67 | 1.485 |
| 30 | 1.99 | 1.74 |
| 35 | 2.00 | 1.745 |

Table 2: Iron Cores in the evolution of Wolf-Rayet stars, with mass loss,[a] from Woosley, Langer, & Weaver (1994). Masses are baryon number masses. The $M_{3500}$ and $M_{4500}$ are the enclosed masses at 3500 and 4500 km, respectively. They were kindly furnished us privately by Stan Woosley.

| Initial He Star Mass/$M_\odot$ | Final He Star Mass/$M_\odot$ | Fe Core Mass/$M_\odot$ | $M_{3500}/M_\odot$ | $M_{4500}/M_\odot$ |
|---|---|---|---|---|
| 5 | 2.82 | 1.38 | 1.55 (1.39) | 1.59 (1.42) |
| 7 | 3.20 | 1.42 | 1.67 (1.485) | 1.71 (1.52) |
| 10 | 3.51 | 1.49 | 1.69 (1.50) | 1.73 (1.53) |
| 20 | 3.55 | 1.49 | 1.70 (1.51) | 1.77 (1.56) |

---

[a] When several cases for a star are given, we have taken only case A. Metallicity 0.02 was considered. Numbers in parentheses are gravitational mass, obtained from the baryon number masses by the binding energy correction of eq (2).

Table 3: The measured masses of 18 compact objects.

| Source | Type[a] | Mass ($M_\odot$)[b] | Reference |
|---|---|---|---|
| SMC X-1 | HMXB | $1.17^{+0.36}_{-0.32}$ | Van Kerkwijk et al. 1995a |
| LMC X-4 | HMXB | $1.47^{+0.44}_{-0.39}$ | Van Kerkwijk et al. 1995a |
| Vela X-1 | HMXB | $1.85^{+0.69}_{-0.47}$ | Van Kerkwijk et al. 1995a |
| Cen X-3 | HMXB | $1.09^{+0.57}_{-0.52}$ | Van Kerkwijk et al. 1995a |
| 1538−522 | HMXB | $1.06^{+0.41}_{-0.34}$ | Van Kerkwijk et al. 1995a |
| Her X-1 | IMXB | $1.47^{+0.23}_{-0.37}$ | Van Kerkwijk et al. 1995a |
| 1700−377 | HMXB | $1.8(4)$[c] | Heap & Corcoran 1992 |
| B1534+12 | BPSR | 1.3378(34) | Arzoumanian 1995 |
| B1534+12c | BPSR | 1.3405(34) | Arzoumanian 1995 |
| J1713+0747 | BPSR | >1.2 | Camilo 1995 |
| B1802−07 | BPSR | 1.28(32) | Arzoumanian 1995 |
| B1855+09[b] | BPSR | $1.50^{+0.52}_{-0.28}$ | Kaspi et al. 1994 |
| B1913+16 | BPSR | 1.442(6) | Taylor & Weisberg 1989 |
| B1913+16c | BPSR | 1.386(6) | Taylor & Weisberg 1989 |
| B2127+11C | BPSR | 1.38(8) | Deich 1995 |
| B2127+11Cc | BPSR | 1.34(8) | Deich 1995 |
| B2303+46 | BPSR | 1.20(52) | Arzoumanian 1995 |
| B2303+46c | BPSR | 1.40(48) | Arzoumanian 1995 |

[a]The abbreviations mean High Mass X-ray Binary, Intermediate Mass X-ray Binary, and Binary Pulsar, respectively. A lowercase c appended to a pulsar name is used to refer to the unseen companion, which is also thought to be a neutron star.

[b]All errors or limits refer to the 95% confidence region. Numbers in parentheses are errors in the last digits. If a one-sigma error was specified in the quoted reference, it was simply doubled. In case of pulsar B1855+09 this is somewhat dubious, because it is the only such case with asymmetric errors. Nonetheless, the confidence contours in the reference show that the limits we quote are roughly correct.

[c]This mass is rather less rigourous and reliable than the others, but it is included because it features in our discussion.

---

# Figure Captions

Fig. 1.— Measured masses of 18 compact objects. X-ray binaries are at the top, radio pulsars and their companions at the bottom. The vertical dashed line indicates our preferred value of $M_{\rm NS,max} = 1.50\,M_\odot$.

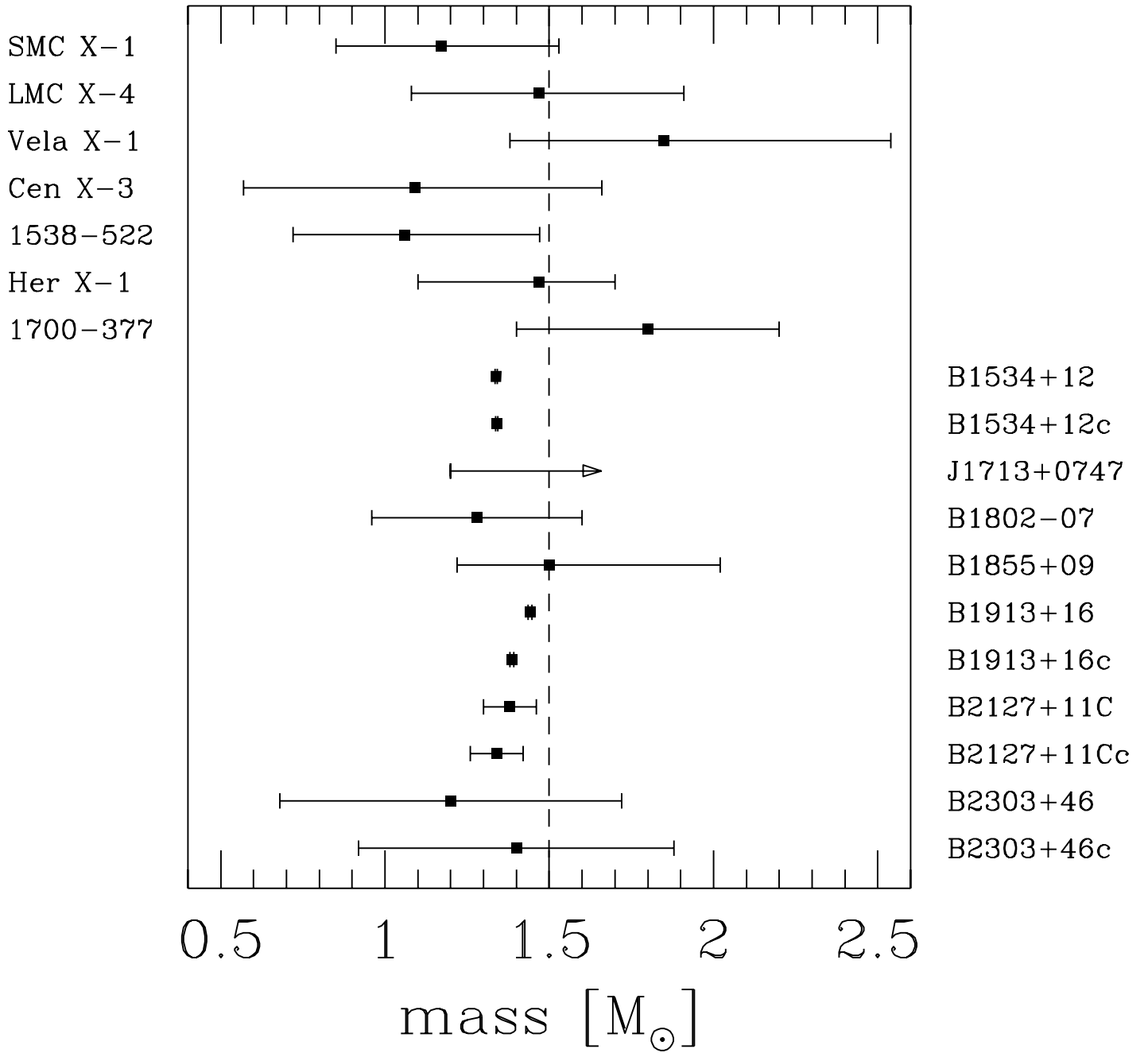